\documentclass[%
 preprint,  
 amsmath,amssymb,
 aps,
]{revtex4}
\usepackage{mathrsfs}
\usepackage{amsfonts}
\usepackage{amsmath}
\usepackage{amssymb}
\usepackage{array}
\usepackage{verbatim}
\usepackage{epsfig}
\usepackage{graphicx}
\usepackage{amsbsy}
\usepackage{bm}
\usepackage[dvipsnames]{xcolor} 
\usepackage{dirtytalk}
\usepackage{csquotes}
\usepackage{graphicx} 
\usepackage[font=small,labelfont=bf]{caption} 
\usepackage[subrefformat=parens]{subcaption}

\usepackage{import}
\input{epsf}
\usepackage{psfrag}
\usepackage{xcolor}
\usepackage{slashed}
\usepackage{hyperref}
\usepackage{ulem}
\usepackage{float}

\begin{document}


\title{Small-$x$ dynamics of the unpolarised color dipole gluon TMD PDFs for all transverse momenta }
\author{Mariyah Siddiqah$^{1}$, Nahid Vasim$^{2}$, Mushood Nabi$^{3}$}

\affiliation{$^{1}$Department of Physics, University of Kashmir, Kupwara campus.}
\affiliation{$^{2}$Indian Institute of Technology Delhi, New Delhi, India}

\affiliation{$^{3}$Department of Physics, University of Kashmir, srinagar.}

\date{\today}

\begin{abstract}
A general solution of the Balitsky–Kovchegov (BK) equation well describes the partonic behavior within the saturation regime and beyond this limit. This solution can potentially define the unpolarized color dipole gluon transverse-momentum-dependent (TMD) distribution in the small $x$ regime for full $k_{\perp}$ range. In this work, we Fourier-Bessel transform this general S-matrix of the BK equation and, within the leading logarithmic approximation, obtain a closed-form expression of unpolarised Gluon TMD. The distribution exhibits a smooth and well-behaved $k_{\perp}$ dependence from low to high
transverse momenta. Numerical evaluation for  different values  of $x$ accessible at the Electron–Ion Collider (EIC), shows a characteristic inversion of
the $x$-ordering at the saturation scale. The result is in close agreement with the behavior observed in the unintegrated color dipole gluon distribution computed within the Bartels–Golec-Biernat–Kowalski (BGK) model, suggesting that this feature is a model-independent signature of gluon
saturation.
\end{abstract}

\maketitle

\section{\label{sec:level1}Introduction}
Understanding the internal structure of hadrons, including nucleons and nuclei, is one of the primary goals of hadron physics. The dynamics of gluons is particularly difficult to understand. At leading power in the hard scale, the internal structure of the proton is described by the parton distribution functions (PDFs)\cite{Alekhin_2017,Bailey_2021}. When we look at the transverse momentum of partons in relation to the proton's motion, this description is extended to Transverse-momentum dependent distributions (TMDs). These distributions capture both the longitudinal momentum fraction $x$ and the intrinsic transverse momentum $k_\perp$ of partons within a fast-moving hadron\cite{Scimemi:2019mlf}. TMDs are defined for both quarks and gluons, and they appear in the expressions for the soft part of cross sections in lepton-hadron and hadron-hadron collisions. While we have made significant advances in understanding quarks from both the theoretical and phenomenological perspectives, the gluon sector remains comparatively less explored. Gluons play a dominant role in governing the behavior of hadronic matter at high energies or when the Bjorken variable $x$ is small. In this kinematic regime, the gluon density increases rapidly, and non-linear QCD effects become more important, driving the transition from the dilute to the dense gluon regime\cite{iancu2002colourglasscondensateintroduction,Iancu_2004}. Investigating  this transition is one of the important topics of interest of current and future high-energy experiments,including the Electron–Ion Collider (EIC)\cite{Goto:2025whi,us_eic_assessment_2018, abdulkhalek_snowmass_2021, burkert_2023, amoroso_snowmass_2022}.

TMDs are sensitive, in particular, to correlations between the parton  transverse momentum and spin, both of the parton itself and the parent hadron. These correlations give insights into the dynamics of partons in the transverse plane in momentum space. Thus, TMDs are comparable and directly complementary to the generalized parton distributions (GPDs)\cite{Diehl_2016}, which encode much of the same physics but in position space rather than momentum space\cite{PhysRevD.104.076028,KAUR201880}. Formally, TMDs access the correlations between a parton orbital angular momentum (OAM) and the  spin of parton or hadron, as TMDs requires contributions from wave function components that carry nonzero orbital angular momentum\cite{BHATTACHARYA2024139134}.

Because their operator definitions include gauge links (wilson lines), the TMDs are not universal. The gluon TMDs at the leading twist are given by parameterizing the gluon-gluon correlation function. For spin-$1/2$ particles at the leading twist, the gluon-gluon correlator can be parameterized into eight types of gluon TMDs\cite{PhysRevD.63.094021,PhysRevD.107.014008,XIE2023137973}. In the small-$x$ limit, where gluon densities grow rapidly and nonlinear saturation effects become significant, these gluon TMDs admit a natural description within the Color Glass Condensate (CGC)\cite{Iancu2001,iancu2002colourglasscondensateintroduction} framework, where they can be expressed in terms of Wilson line correlators and studied through evolution equations\cite{XIAO2017104}. The gluon TMDs, when expressed in Wilson line operators, can also be reduced as a function of  dipole 
$S$-matrix. This connection between the TMD framework and the CGC not only provides a systematic way to compute gluon TMDs at small $x$
 from first principles, but also links the transverse momentum structure of the gluon distribution directly to the nonlinear evolution of the dipole amplitude governed by the BK equation\cite{Balitsky_1996,PhysRevD.83.105005}. In particular, the unintegrated color dipole gluon distribution, which constitutes the central focus of the present work, is given by

\begin{widetext}
\begin{eqnarray}
  x\mathcal{F}^{(1)}(x,k_{\perp}) =\frac{S_{\bot}N_c}{2\pi^2\alpha_{s}}k_\perp^2 
 \int\frac{d^2r_{\bot}}{(2\pi)^2}e^{-ik_{\bot}.r_{\bot}} 
 \frac{1}{N_c}<{\rm{Tr}} [U(x_{\perp},Y)U^{\dagger}(y_{\perp},Y)]>_x.\nonumber\\
\label{GDP} 
\end{eqnarray}
\end{widetext}
The quantity in the angle brackets is precisely the forward dipole
S-matrix $S(x, r_\perp)$, encoding the amplitude for a color-neutral quark-antiquark pair of transverse size $r_\perp = x_\perp - y_\perp$ to scatter on the target without interaction. The nonlinear small-$x$
 evolution of this
$S$-matrix is governed by the Balitsky-Kovchegov (BK) equation in the large-$N_c$ limit, which incorporates quantum unitarity corrections that suppress the otherwise unbounded growth of gluon densities at small-$x$\cite{Balitsky_1996,PhysRevD.61.074018}.

Despite the physical importance of the dipole gluon distribution, a complete analytical description that is valid uniformly across the full transverse momentum spectrum remains an open problem. Existing analytical results are strictly piecewise. Different kinematic regimes of $k_\perp$ probe different regimes of $r_\perp$, each of which admits its own approximate analytical solution of the BK equation. In the perturbative tail $k_\perp \gg Q_s$, the distribution falls as $1/k_\perp^2$, consistent with perturbative QCD expectations \cite{Cai_2023,kutak2026balitsky}. Near the saturation boundary $k_\perp \gtrsim Q_s$, where the $S$-matrix takes an anomalous dimension form, the UDGD exhibits a power-law behavior  $(Q_s^2/k_\perp^2)^{\gamma_{cr}}$ with the critical anomalous dimension $\gamma_{cr} \approx 0.63$ \cite{IANCU2002327}. Deep inside the saturation region, $k_\perp \ll Q_s$, the unintegrated dipole gluon distribution was derived in Ref.~\cite{Siddiqah_2018} using the Levin--Tuchin solution of the BK equation~\cite{Levin:1999mw}, yielding the compact expression
\begin{equation}
x\mathcal{F}^{(1)}(x,k_\perp)\Big|_{Q_s \gtrsim k_\perp \gg \Lambda_{QCD}} \approx -\frac{S_\perp N_c \tau}{\pi^3 \alpha_s} \ln\frac{k_\perp^2}{4Q_s^2(Y)} \exp\!\left(-\tau \ln^2\frac{k_\perp^2}{4Q_s^2(Y)}\right),
\label{LT_result}
\end{equation}
where $\tau \approx 0.2$ is a constant arising from the saddle-point condition along the saturation line. This Sudakov-like double-logarithmic structure within the saturation region is a significant result, implying connections between the nonlinear small-$x$ evolution and the soft-gluon resummation physics encoded in the Collins-Soper-Sterman (CSS) framework \cite{Collins:1984kg,XIAO2017104}. The above result motivates the search for a unified analytical expression for $x\mathcal{F}^{(1)}(x, k_\perp)$ valid across the \textit{entire} transverse momentum spectrum ,smoothly interpolating from the deeply saturated regime $k_\perp \ll Q_s$ through the saturation boundary $k_\perp \sim Q_s$ and into the perturbative tail $k_\perp \gg Q_s$ is still lacking.  
The unintegrated color dipole gluon distribution is directly accessible across the full $k_\perp$ spectrum in several experimentally relevant processes. In inclusive and diffractive deep inelastic scattering (DIS) in $ep$ and $eA$ collisions, the structure functions $F_2$ and $F_L$ are sensitive to the dipole amplitude across the saturation boundary~\cite{LUSZCZAK2022137582}. In $pA$ collisions, direct photon-jet azimuthal correlations probe the color dipole TMD at small $x$ across the full transverse momentum range~\cite{Dominguez:2011wm}, while forward dijet and di-hadron production are governed by the Fourier transform of the dipole amplitude from the deeply saturated to the perturbative regime~\cite{Marquet:2016cgx, BGK}. At the EIC, semi-inclusive deep inelastic scattering (SIDIS) at small $x$ provides perhaps the cleanest environment for extracting the color dipole gluon TMD across all three kinematic regions — deep saturation, saturation boundary, and perturbative tail - making a unified analytic expression such as Eq.~\eqref{final} directly relevant to these studies.

In the present work, we derive a color dipole distribution for all $k_{\perp}$ regions. We employ a solution of the linearized BK equation \cite{Siddiqah_2017}, which, unlike previous solutions, is valid both deep inside and well outside the saturation region. Using this solution, we analytically derive a compact expression for the color-dipole gluon distribution. We believe this result constitutes a timely and important theoretical contribution to the emerging precision program of small-$x$ gluon tomography at the EIC.

The paper is organized as follows. In Sec .~II, we introduce the general $S$-matrix ansatz, derive the color dipole gluon TMD via the Fourier-Bessel transform and evaluate finite contributions to get the final closed-form result in the leading-logarithm approximation. We find that the divergent part vanishes. Section~III presents a detailed numerical analysis of the distribution as a function of $k_\perp/Q_s$ and $x$, including a comparison with the BGK model. We conclude with an outlook in Sec.~IV. Technical details of the Fourier transform are shown in Appendix~A.

\section{\label{sec:level2} Color Dipole Gluon TMD from a General S-Matrix Ansatz}

The scattering of a color dipole off a large hadronic target is most naturally described 
in terms of the forward  $S$-matrix, which encodes the probability amplitude for 
the dipole to pass through the target without inelastic interaction. In the eikonal 
approximation, the $S$-matrix is expressed as a two-point correlator of light-like Wilson 
lines
\begin{eqnarray}
    S(\mathbf{r}_{\perp},Y) = \frac{1}{N_c}
    \Big\langle \mathrm{Tr}\big[W(\mathbf{x}_{\perp},Y)\,
    W^{\dagger}(\mathbf{y}_{\perp},Y)\big]\Big\rangle,
    \label{eq:Smat_def}
\end{eqnarray}
where $\mathbf{r}_{\perp} = \mathbf{x}_{\perp} - \mathbf{y}_{\perp}$ denotes the 
transverse dipole separation and $Y$ is the rapidity variable. The rapidity evolution 
of $S(\mathbf{r}_{\perp},Y)$ is governed by the BK 
equation, which resums the leading 
$\ln(1/x)$ contributions and incorporates unitarity corrections due to gluon 
saturation. Despite its well-established status,
analytic solutions to the BK equation are only known in specific kinematic
limits. At moderately small dipole sizes, the McLerran–Venugopalan (MV) 
model~\cite{McLerran:1993ni,McLerran:1993ka,McLerran:1994vd} provides the natural 
initial condition for the evolution; it yields an $S$-matrix that is Gaussian in the 
dimensionless scaling variable $r_{\perp}Q_s(Y)$. Deep inside the saturation region, 
where $r_{\perp}Q_s(Y) \gg 1$, the appropriate asymptotic form is instead given by 
the Levin–Tuchin (LT) solution~\cite{Levin:1999mw,Levin:2000mv}, which is Gaussian 
in the logarithm of $r_{\perp}Q_s(Y)$. Although each solution accurately captures 
the physics in its respective kinematic domain, neither is capable of providing a 
unified description that smoothly interpolates between the dilute and dense regimes of QCD matter. To overcome this limitation, a general analytic solution to the BK equation valid 
over the full kinematic range of saturation dynamics was derived in 
Ref.~\cite{Siddiqah_2017}. Expressed in terms of the dilogarithm (Spence's function) 
$\mathrm{Li}_2$, this solution reads
\begin{eqnarray}
    S = S_0 \exp\!\Big(\tau\;\mathrm{Li}_2\!\big(-\lambda\, r_{\perp}^2 Q_s^2\big)\Big),
    \label{eq:genSmat}
\end{eqnarray}
where the parameters $\tau = 1/4.88$ and $\lambda = 7.22$ are fixed by matching to 
known limiting solutions, and $S_0$ is an overall normalization constant independent 
of the initial conditions. One may verify that Eq.~\eqref{eq:genSmat} correctly 
reproduces both limiting forms in the appropriate regimes. The two asymptotic limits of the dipole (S)-matrix can be obtained directly from Eq.~\eqref{eq:genSmat}. In the dilute regime, $(r_{\perp}Q_s \ll 1/\lambda)$, the expression reduces to a Gaussian form in the scaling variable       $  (\tau=r_{\perp}Q_s(Y))       $, consistent with the behavior of the McLerran–Venugopalan model and Golec-Biernat–Wüsthoff model up to a model-dependent variance. While as,  in the black-disc limit, $(r_{\perp}Q_s \gg \lambda)$, this expression reproduces the Levin–Tuchin solution characterized by the quadratic logarithmic dependence in the exponent.


Having established the general analytic form of the dipole $S$-matrix, we are now in a position to derive the unintegrated color dipole gluon TMD.  Substituting the general analytic representation of the $S$-matrix, Eq.~\eqref{eq:genSmat}, into Eq.~\eqref{GDP}, and making use of the identity of the dilogarithmic function, yields the explicit integral expression

\begin{eqnarray}
x\mathcal{F}^{(1)}(x,k_{\perp}) \;=\; \frac{S_{\perp}\,N_c}{2\pi^2\alpha_{s}}\,k_\perp^2\; \int\!\frac{d^2\mathbf{r}_{\perp}}{(2\pi)^2}\, e^{-i\mathbf{k}_{\perp}\cdot\mathbf{r}_{\perp}}\, \exp\!\left( -\tau\left[ \frac{1}{2}\ln^2\!\big(\lambda\,r_{\perp}^2\,Q_s^2\big) + \mathrm{Li}_2\!\left(\frac{-1}{\lambda\,r_{\perp}^2\,Q_s^2}\right) \right] \right). \label{eq:xG_integral}
\end{eqnarray}

The two-dimensional Fourier transform in Eq.~\eqref{eq:xG_integral} is evaluated following the method of Ref.~\cite{Siddiqah_2018}; full technical details are presented in Appendix~\ref{FT}. The resulting expression, given in Eq.~\eqref{finaleq},takes the form of a ratio of Gamma functions encoding the complete analytic structure of the distribution. This ratio is then systematically expanded using Bell polynomials, as detailed in Eq.~\eqref{gammaf}, furnishing a compact analytic representation of the higher-order contributions. The  Taylor series  expansion of the full expression leads the only potential source of ultraviolet divergence as $1/\eta$
 pole. We therefore decompose the distribution as
\begin{eqnarray}
   x\mathcal{F}^{(1)}(x,k_{\perp}) = 
    xf^{(1)}_{g}\big|_{\rm div} + xf^{(1)}_{g}\big|_{\rm fin}.
\end{eqnarray}
and treat each contribution in turn. 
The divergent part reads
\begin{eqnarray}
xf^{(1)}_{g}\big|_{\rm div}=\frac{S_{\perp}N_c}{2\pi^3\alpha_s}\lim_{\eta\to0}\sum_{n=0}^{\infty}\frac{(-\tau)^n}{2^n n!}\frac{\partial^{2n}}{\partial\eta^{2n}}\left[\frac{1}{\eta}\left(\frac{k_\perp^2}{4\lambda Q_s^2}\right)^\eta\sum_{j=1}^{\infty}\frac{B_j\!\left(-\tau/1^2,-2!\tau/2^2,\ldots\right)}{j!(j-1)!^2}\left(\frac{k_\perp^2}{4\lambda Q_s^2}\right)^j\right].
\label{eq:div_part}
\end{eqnarray}

The divergent piece, upon applying the Leibniz rule for the $\eta$-derivative and resumming the 
$j$-series via the exponential generating function of Bell polynomials, takes the form
\begin{eqnarray}
    xf^{(1)}_{g}\big|_{\rm div} &=& 
    \frac{S_{\perp}N_c}{2\pi^3\alpha_{s}}
    \lim_{\eta\to 0}
    \left(\frac{k_{\perp}^2}{4\lambda Q_s^2}\right)^{\!\eta}
    \frac{1}{\eta}\,
    \exp\!\left[\frac{-\tau}{2\eta^2}\right]
    \exp\!\left[-\tau\,\widetilde{\mathrm{Li}}_2\!\left(
    \frac{k_{\perp}^2}{4\lambda Q_s^2}\right)\right]
    \,
    \nonumber\\
    \label{eq:div_vanishes}
\end{eqnarray}
where $ \widetilde{\mathrm{Li}}_2(z) = \sum_{k=1}^{\infty}
    \frac{z^k}{k^2\,(k-1)!^{2/k}}$ is a modified dilogarithm. The Gaussian suppression factor $\exp(-\tau/2\eta^2)$ vanishes faster than any power of $\eta$
as $\eta \to 0$, so the divergent contribution vanishes identically. This confirms the absence of any genuine ultraviolet divergence in the color dipole gluon TMD and establishes the internal consistency of the regularization scheme.


The finite part of the gluon TMD is given by
\begin{eqnarray}
    xf^{(1)}_{g}\big|_{\rm fin} &=& 
    \frac{S_{\perp}N_c}{2\pi^3\alpha_{s}}
    \lim_{\eta\to 0}\sum_{n=0}^{\infty}
    \frac{\partial^{2n}}{\partial\eta^{2n}}
    \frac{(-\tau)^n}{2^n\,n!}
    \left(\frac{k_{\perp}^2}{4\lambda Q_s^2}\right)^{\!\eta}
    \sum_{j=1}^{\infty}
    \frac{B_j\!\left(-\tau/1^2,\,-2!\tau/2^2,\,\ldots\right)}
    {j!\,(j-1)!^2}
    \left(\frac{k_{\perp}^2}{4\lambda Q_s^2}\right)^{\!j}
    \nonumber\\
    &\times&\sum_{m=0}^{\infty}
    \frac{(-1)^m}{m!}
    B_{m+1}\!\left(2\psi^{(0)}(j),\,-2\psi^{(1)}(j)+\tfrac{\pi^2}{3},
    \,2\psi^{(2)}(j),\,\ldots\right)\eta^m,
    \label{eq:fin_part}
\end{eqnarray}
where $\psi^{(n)}(j)$ denotes the polygamma function of order $n$. Applying the generalized Leibniz rule for the $\eta$
derivative along with the convolution property of Bell polynomials, and taking the limit 
$\eta\to 0$ yields
\begin{eqnarray}
    &&\lim_{\eta\to 0}\frac{\partial^{2n}}{\partial\eta^{2n}}
    \left[\left(\frac{k_{\perp}^2}{4\lambda Q_s^2}\right)^{\!\eta}
    \sum_{m=0}^{\infty}\frac{1}{m!}
    B_m\!\left(-2\psi^{(0)}(j),\,2\psi^{(1)}(j)-\tfrac{\pi^2}{3},\,
    -2\psi^{(2)}(j),\,\ldots\right)\eta^m
    \right]
    \nonumber\\
    &=& B_{2n}\!\left(\ln\frac{k_{\perp}^2}{4\lambda Q_s^2}-2\psi^{(0)}(j),\;
    2\psi^{(1)}(j)-\tfrac{\pi^2}{3},\;2\psi^{(2)}(j),\,\ldots\right),
    \label{eq:derivative_result}
\end{eqnarray}

\medskip
In the approximation where $\ln(k_\perp^2/4\lambda Q_s^2)$
 dominates over the polygamma corrections and the Bell polynomials are retained to leading order,  the finite contribution for $j\geq 1$
 simplifies to;

\begin{eqnarray}
    xf^{(1)}_{g}\big|_{\rm fin}^{j\geq 1} &=&
   \frac{S_{\perp}N_c}{\pi^3\alpha_{s}}
    \exp\!\left[-\frac{\tau}{2}\ln^2\!\left(\frac{k_{\perp}^2}{4\lambda Q_s^2}
    \right)\right]
    \left(\frac{\tau\gamma_E k_{\perp}^2}{4\lambda Q_s^2}\right).
    \label{eq:fin_jgeq1_closed}
\end{eqnarray}
where $\gamma_E \approx 0.5772$ is the Euler–Mascheroni constant, which enters through the digamma function. The $j = 0$
 contribution, in the leading-logarithm (LL) approximation, reduces to a pure Gaussian

\begin{eqnarray}
    xf^{(1)}_{g}\big|_{j=0} &=&
    \frac{S_{\perp}N_c}{2\pi^3\alpha_{s}}
    \exp\!\left[-\frac{\tau}{2}\ln^2\!\left(\frac{k_{\perp}^2}{4\lambda Q_s^2}
    \right)\right].
    \label{eq:j0_LL}
\end{eqnarray}
Combining Eqs.~\eqref{eq:fin_jgeq1_closed} and~\eqref{eq:j0_LL}, the complete color dipole gluon TMD in the leading-logarithm approximation is
\begin{eqnarray}
x\mathcal{F}^{(1)}(x,k_{\perp}) = \frac{S_{\perp}N_c}{2\pi^3\alpha_{s}}
    \exp\!\left[-\frac{\tau}{2}\ln^2\!\left(\frac{k_{\perp}^2}{4\lambda Q_s^2}
    \right)\right]
    \left[1 + \frac{\tau\gamma_E k_{\perp}^2}{2\lambda Q_s^2}\right].
    \label{final}\end{eqnarray}

Equation~\eqref{final} is the principal analytic result of this
work. This result is a unified, closed-form expression for the unintegrated color dipole gluon TMD valid across the full transverse momentum spectrum. We now study its behavior in each of the three physically distinct kinematic regions, showing that it correctly reproduces the known limiting forms and encodes the full transition between them. Deep inside the saturation region, $k_\perp \ll Q_s$, the linear correction $\tau\gamma_E k_\perp^2/2\lambda Q_s^2 \ll 1$ is negligible and the distribution becomes the pure log-Gaussian,
which suppresses the distribution faster than any power law in the limit $k_\perp \to 0$. This reproduces exactly the result of Ref.~\cite{Siddiqah_2018} derived from the Levin--Tuchin solution~\cite{Levin:1999mw, Levin:2000mv}, confirming the internal consistency of the derivation. The overall prefactor $S_\perp N_c/2\pi^3\alpha_s \sim \mathcal{O}(1/\alpha_s)$ reflects the characteristic scaling of the gluon occupation number in the saturation regime. The double-logarithmic suppression structure is similar of the Sudakov factor in the Collins--Soper--Sterman (CSS) framework, suggesting a deep connection between nonlinear small-$x$ evolution and that of soft-gluon resummation physics.

At the saturation boundary $k_\perp \sim Q_s$, the saturation scale marks the transverse momentum at which the gluon occupation number reaches $\mathcal{O}(1/\alpha_s)$, signalling the onset of the nonlinear BK regime and acting as the phase boundary between the dense and dilute regimes of QCD matter. In this region, both the log-Gaussian factor and the linear correction contribute comparably, and the full structure of Eq.~\eqref{final} is required, with the distribution proportional to $\mathcal{O}(1/\alpha_s)$ in direct accordance with the \enquote{saturon} criterion. 

In the dilute regime, $k_\perp \gg Q_s$, the linear correction dominates and the distribution 
exhibits the perturbative $k_\perp^2/Q_s^2$ rise consistent with the MV and GBW models and with the large-$k_\perp$ behavior from JIMWLK evolution. The log-Gaussian envelope guarantees the normalizability by suppressing the distribution faster than any power of $k_\perp$ as $k_\perp\to\infty$. Moreover, a unifying feature of Eq.~\eqref{final} across all three regions is that the distribution depends on $k_\perp$ and $Q_s$ exclusively through the ratio $k_\perp^2/Q_s^2$ and therefore exhibits geometric scaling.

\section{Numerical Results and Discussion}\label{numerics}

The analytic result Eq.~\eqref{final}, derived in Sec.~\ref{sec:level2}, constitutes main result of this study. After discussing its analytic structure and asymptotic limits in the preceding section, we now proceed to a numerical study of the distribution and its phenomenological consequences. We first examine the overall shape of the distribution as a function of $k_\perp/Q_s$, then study its $x$-dependence for three representative values accessible at the EIC.

\begin{figure}[H]
    \centering
    \includegraphics[width=15cm]{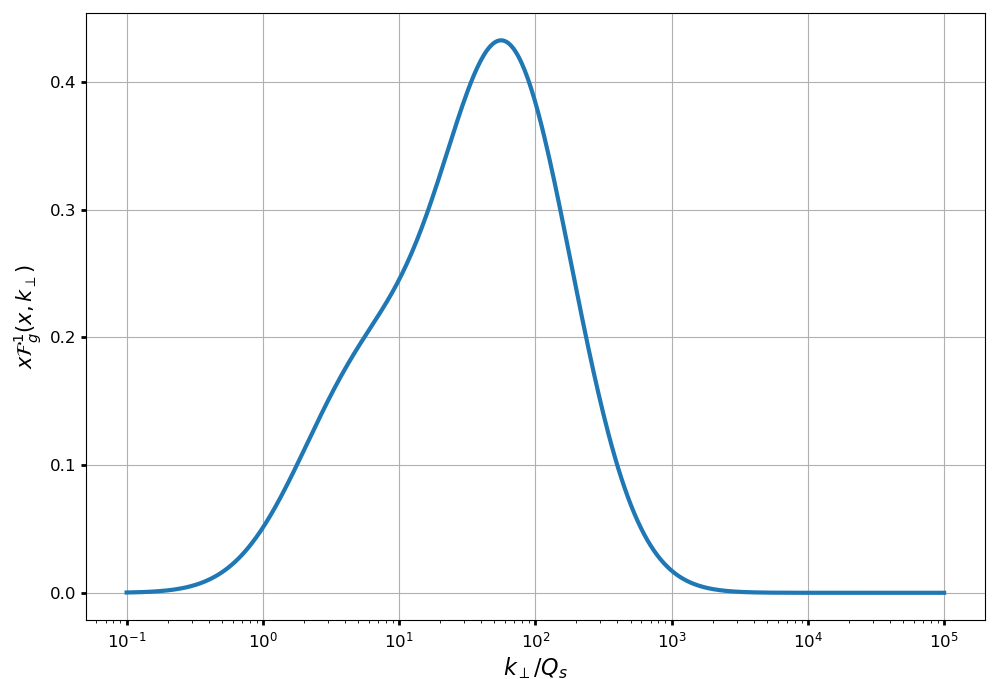}
    \caption{The color dipole gluon TMD $x\mathcal{F}^{(1)}(x,k_{\perp})$
             as a function of $k_\perp/Q_s$, evaluated from the analytic
             result Eq.~\eqref{final}.}
    \label{fig:scaling}
\end{figure}

In Fig.~\ref{fig:scaling}, we plot $x\mathcal{F}^{(1)}(x,k_\perp)$ as a function of the dimensionless ratio $k_\perp/Q_s$. Expressing the result in terms of this scaling variable makes the saturation structure manifest. It reveals the geometric scaling property of the distribution established analytically in Sec.~\ref{sec:level2}: the TMD depends on $k_\perp$ and $Q_s$ only through their ratio, a robust prediction of saturation physics confirmed numerically for all TMD gluon distributions. At small $k_\perp/Q_s$, the double-logarithmic Gaussian factor in Eq.~\eqref{final} drives the distribution to zero, providing a Sudakov-type suppression at low transverse momenta consistent with $S$-matrix unitarity. The distribution then rises, reaches a well-defined peak and falls off smoothly and monotonically at large $k_\perp/Q_s$, the log-Gaussian envelope ensuring normalizability across the entire range. This smooth, peak-shaped behavior is a direct consequence of the global validity of the general $S$-matrix ansatz of Ref.~\cite{Siddiqah_2017}, which interpolates continuously between the dense and dilute regimes without any piecewise matching, and could not have been obtained from the previously available piecewise analytic results.

\begin{figure}[H]
    \centering
    \includegraphics[width=15cm]{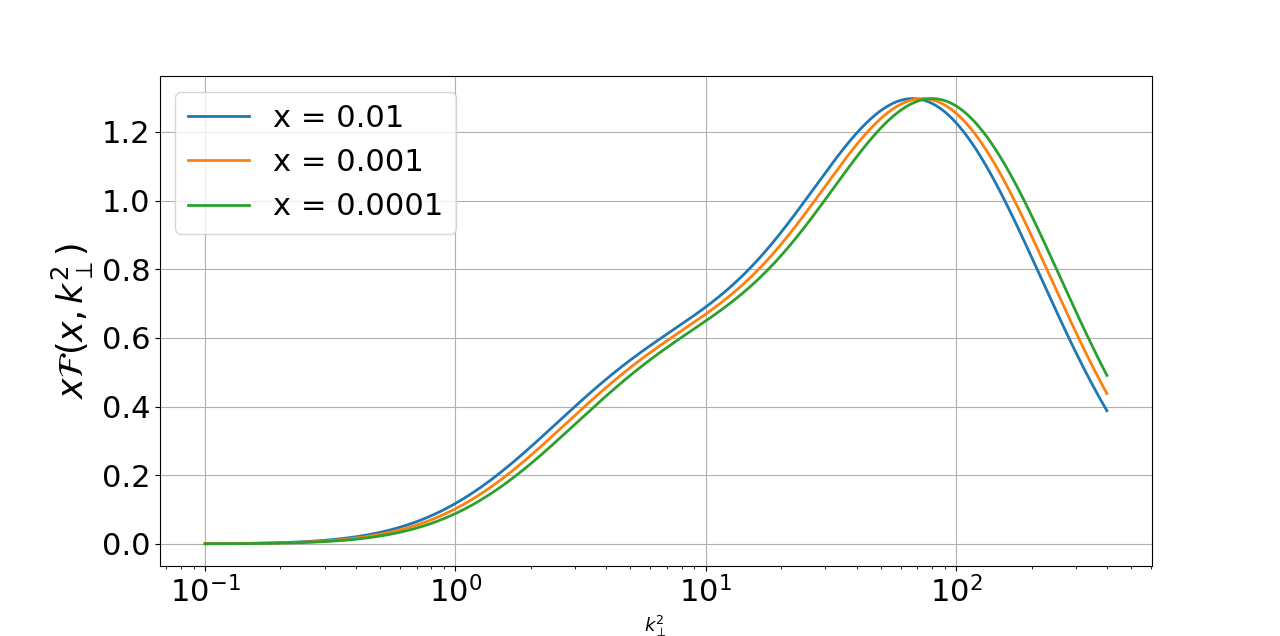}
    \caption{Color dipole gluon TMD $x\mathcal{F}^{(1)}(x,k_{\perp}^2)$
             as a function of $k_{\perp}^2$ for $x = 0.01$ (blue),
             $0.001$ (orange), and $0.0001$ (green). The saturation scale evolves with
             $x$ according to Eq.~\eqref{eq:Qsevol} with
        $Q_{s0}=1\,\mathrm{GeV}$.}
    \label{fig:3}
\end{figure}

In Fig.~\ref{fig:3}, we evaluate Eq.~\eqref{final} as a function of $k_\perp^2$ for three representative values of $x$ accessible at the EIC: $x = 0.01$, $0.001$, and $0.0001$. The saturation scale is taken to evolve with $x$ according to
\begin{eqnarray}
Q_s(x) = Q_{s0}\exp\!\left(\frac{2.44\,\alpha_s}{2\pi}\ln\frac{1}{x}\right), \label{eq:Qsevol}
\end{eqnarray}

with $Q_{s0} = 1~\text{GeV}$, reflecting the leading-order BFKL growth of the saturation scale with decreasing $x$~\cite{Levin:1999mw}. In the limit $k_\perp^2 \to 0$, the distribution goes to zero for all three curves, independently of $x$, which is a direct consequence of the double-logarithmic suppression in Eq.~\eqref{final}. This behavior is directly analogous to Sudakov suppression in perturbative QCD. As $k_\perp^2$ increases beyond the respective saturation scales, the distribution starts to depend on  $x$. In the pre-saturation region $k_\perp \lesssim Q_s(x)$, the distribution is larger for higher values of $x$, reflecting the correspondingly smaller saturation scale and wider transverse-momentum profile. Beyond the saturation peak, however, the hierarchy inverts: lower-$x$ distributions overtake those at higher $x$ and continue to grow more steeply at large $k_\perp^2$. This crossing behavior is a characteristic signature of gluon saturation dynamics and has been observed in related unintegrated gluon distributions computed within the Bartels--Golec-Biernat--Kowalski (BGK) framework~\cite{BGK}, establishing it as a robust, model-independent prediction of small-$x$ saturation physics rather than an artefact of the particular ansatz employed here.
 
Taken together, the results of this section establish that Eq.~\eqref{final} correctly and quantitatively captures the physics of the unintegrated color dipole gluon TMD across all kinematic regimes relevant to the EIC.

\section{Conclusion}

In this work we have derived a unified, closed-form analytic expression for the unintegrated color dipole gluon TMD, valid across the full transverse momentum spectrum. The analytic expression that smoothly interpolates from the deeply saturated regime $k_\perp \ll Q_s$ across the saturation boundary $k_\perp \sim Q_s$ and into the perturbative tail $k_\perp \gg Q_s$  has remained an open problem despite the physical importance of this distribution for small-$x$ Proton tomography at the EIC and for cross-section calculations in $pA$ collisions. To address this issue, we employed the general analytic solution of the BK equation of Ref.~\cite{Siddiqah_2017}, expressed in terms of the dilogarithm $\mathrm{Li}_2$ and valid across the full kinematic range of saturation dynamics. By substituting this $S$-matrix into the defining relation for the color dipole gluon distribution, we obtained an explicit integral expression whose two-dimensional Fourier-Bessel transform was evaluated analytically. The resulting Gamma function ratio was systematically expanded using Bell polynomials, and the ultraviolet-divergent contribution was shown to vanish identically, confirming the internal consistency of the regularization scheme, thus leaving a finite expression that was resummed in the leading-logarithm approximation to yield the principal result of this work, Eq.~\eqref{final}.

The analytic structure of Eq.~\eqref{final} encodes physically significant information. The leading log-Gaussian factor $\exp[-\frac{\tau}{2}\ln^2(k_\perp^2/4\lambda Q_s^2)]$ originates directly from the nonlinear BK evolution and reproduces exactly the result of Ref.~\cite{Siddiqah:2018qey} derived from the Levin Tuchin solution~\cite{Levin:1999mw, Levin:2000mv} in the deep saturation limit, confirming the self-consistency of the derivation. The perturbative correction $[1 + \tau\gamma_E k_\perp^2/2\lambda Q_s^2]$, governs the smooth transition from the saturated to the dilute regime. The distribution satisfies geometric scaling exactly, depending on $k_\perp$ and $Q_s$ only through $k_\perp^2/Q_s^2$, a signature of all the TMDs, and the overall $\mathcal{O}(1/\alpha_s)$ prefactor is consistent with the saturon criterion at $k_\perp \sim Q_s$. Furthermore, the double-logarithmic Gaussian suppression is structurally parallel to the Sudakov factor in the Collins-Soper-Sterman (CSS) framework~\cite{Collins:1984kg, Collins:1988ig}, pointing to a deep connection between nonlinear small-$x$ evolution and soft-gluon resummation physics~\cite{Xiao:2013yv, Zhou:2017xtb} that merits systematic investigation beyond the leading logarithm level.

The numerical results not only confirm the analytic structure of Eq.~\eqref{final} but also highlight its phenomenological implications. The distribution exhibits a smooth, well-defined peak, vanishes with double-logarithmic suppression as $k_\perp \to 0$, and falls off monotonically in the perturbative tail, all from a single unified expression without piecewise matching. The $x$-dependence, studied for three values $x = 0.01$, $0.001$, and $0.0001$ relevant to the EIC kinematic regime, reveals a characteristic crossing behavior: in the pre-saturation region, the distribution is larger for higher $x$, while beyond the saturation peak, the hierarchy inverts and lower-$x$ distributions grow more steeply. This crossing is a robust, model-independent signature of gluon saturation dynamics, consistent with results obtained within the BGK framework~\cite{BGK}, and its observation in the present analytic result strengthens confidence in the physical reliability of Eq.~\eqref{final}.

From the phenomenological perspective, Eq.~\eqref{final} is directly applicable to a number of processes that probe the color dipole gluon distribution across the full $k_\perp$ spectrum, including inclusive and diffractive DIS structure functions in $ep$ and $eA$ collisions, direct photon-jet correlations in $pA$ collisions, forward dijet and di-hadron production in $pA$ and $eA$ collisions, and SIDIS at the EIC~\cite{Dominguez:2011wm, Marquet:2016cgx, Boroun:2023prd}.In all of these processes, contributions from gluons, particularly near the saturation boundary $k_\perp \sim Q_s$ are significantly important. The unified expression derived here eliminates the systematic uncertainties inherent in piecewise matching procedures. This will allow more reliable extractions of the saturation scale $Q_s$ and other nonperturbative parameters from future EIC data. Several natural extensions of the present work suggest themselves: a systematic comparison of Eq.~\eqref{final} with existing parametrizations including the GBW, MV-IR, and bSat models; application to specific EIC observables; extension of the present derivation to the Weizsäcker--Williams gluon distribution using the same $S$-matrix ansatz; and a systematic investigation of the CSS--Sudakov connection identified here beyond the leading-logarithm approximation. These directions are left for future work.

\appendix
 

\newpage

\section{Fourier Transform of Dipole TMD\label{FT}}

The small-$x$ definition of the colour dipole gluon distribution in transverse-momentum space is:
\begin{eqnarray}
x\mathcal{F}^{(1)}_{g}(x,k_{\perp}) = \frac{S_{\perp}N_c}{2\pi^2\alpha_{s}}\,k_\perp^2 \int\!\frac{d^2r_{\perp}}{(2\pi)^2}\, e^{-i\mathbf{k}_{\perp}\cdot\mathbf{r}_{\perp}}\, S(x,\mathbf{r}_{\perp})\label{eq:app_start}
\end{eqnarray}

with the general $S$-matrix of Ref.~\cite{Siddiqah_2017},
\begin{eqnarray}
S \;=\; S_0\, \exp\!\Big[\tau\,\mathrm{Li}_2\!\big(-\lambda\,r_{\perp}^2\,Q_s^2\big)\Big], \label{eq:app_genS}
\end{eqnarray}

where $\tau = 1/4.88$ and $\lambda = 7.22$ are fixed by matching to the known limiting solutions of the BK equation, and $S_0$ is a normalization constant independent of the initial conditions.Substituting \ref{eq:app_genS} in \ref{eq:app_start} and using dilogrithic reflection  identity we can write

\begin{eqnarray}
x\mathcal{F}^{(1)}_{g}(x,k_{\perp}) = \frac{S_{\perp}N_c}{2\pi^2\alpha_{s}}\,k_\perp^2 \int\!\frac{d^2r_{\perp}}{(2\pi)^2}\, e^{-i\mathbf{k}_{\perp}\cdot\mathbf{r}_{\perp}} \left( \sum_{n=0}^{\infty} \frac{(-\tau/2)^n}{n!} \ln^{2n}\!\!\left[\frac{1}{\lambda r_{\perp}^2 Q_s^2}\right] \right)
\nonumber\\
\times \left( \sum_{j=0}^{\infty} \frac{B_j\!\left(-\tau\dfrac{1!}{1^2},\,-\tau\dfrac{2!}{2^2},\, \ldots,\,-\tau\dfrac{j!}{j^2}\right)}{j!} \left(\frac{-1}{\lambda r_{\perp}^2 Q_s^2}\right)^{\!j} \right), \label{eq:app_expanded}
\end{eqnarray}

where the exponential of the $\mathrm{Li}_2$ series has been expressed using the exponential generating function of complete Bell polynomials $B_j$
\begin{eqnarray}
\exp\!\left(\sum_{k=1}^{\infty} x_k \frac{t^k}{k!}\right) = \sum_{j=0}^{\infty} B_j(x_1, x_2, \ldots)\frac{t^j}{j!},
\end{eqnarray}

with the identification $x_k = -\tau\, k!/k^2$. Treating the $j=0$ term seperately,the contribution of the general $S$-matrix spanning the full kinematic domain enters from $j \geq 1$. We now  expand the exponential to express it in the form of a series, where the n-th
term having $2n$-th derivatives of the dummy variable $\eta$ as

\begin{eqnarray}
x\mathcal{F}^{(1)}_{g}(x,k_{\perp}) = \frac{S_{\perp}N_c}{2\pi^2\alpha_{s}}\,k_\perp^2 \sum_{n=0}^{\infty} \frac{(-\tau)^n}{2^n\,n!} \lim_{\eta\to 0} \frac{\partial^{2n}}{\partial\eta^{2n}} \sum_{j=1}^{\infty} (-1)^j \frac{B_j\!\left(-\tau\dfrac{1}{1^2},\,-\tau\dfrac{2!}{2^2},\,\ldots\right)}{j!}   \nonumber\\
\times\left(\frac{1}{\lambda Q_s^2}\right)^{\!\eta+j} \frac{1}{2\pi} \int_0^{\infty}\!dr_{\perp}\,r_{\perp}\, J_0(r_{\perp}k_{\perp})\,r_{\perp}^{-2(\eta+j)}, \label{eq:app_polar}
\end{eqnarray}
 The  integral now can be  evaluated using the standard Bessel-function identity~\cite{GradshteynRyzhik},
\begin{eqnarray}
  \int_0^{\infty}\!dr\;r^{\mu-1}\,J_0(kr) \;=\; 2^{\mu-1}\,k^{-\mu}\, \frac{\Gamma(\mu/2)}{\Gamma(1-\mu/2)}, \label{eq:app_Bessel}  
\end{eqnarray}

 After the integration our result takes the form
\begin{eqnarray}
x\mathcal{F}^{(1)}_{g}(x,k_{\perp}) = \frac{S_{\perp}N_c}{\pi^3\alpha_{s}} \lim_{\eta\to 0} \sum_{n=0}^{\infty}\sum_{j=1}^{\infty} \frac{\partial^{2n}}{\partial\eta^{2n}} \frac{(-1)^{n+j}\,\tau^n}{2^n\,n!} \frac{B_j\!\left(-\tau\dfrac{1}{1^2},\,-\tau\dfrac{2!}{2^2},\,\ldots\right)}{j!} \left(\frac{k_{\perp}^2}{4\lambda Q_s^2}\right)^{\!\eta+j} \frac{\Gamma(1-\eta-j)}{\Gamma(\eta+j)}. \label{finaleq} \nonumber\\
\end{eqnarray}

\subsection*{Expansion of the Gamma-function ratio}

Expanding the ratio $\Gamma(1-\eta-j)/\Gamma(\eta+j)$ about $\eta = 0$. A systematic Taylor expansion yields
\begin{eqnarray}
\frac{\Gamma(1-\eta-j)}{\Gamma(\eta+j)} \;=\; \frac{(-1)^j}{\left[(j-1)!\right]^2} \left[ \frac{1}{\eta} + \mathcal{A}_0(j) + \mathcal{A}_1(j)\,\eta + \mathcal{A}_2(j)\,\eta^2 + \cdots \right], \label{eq:app_Gamma_expand}
\end{eqnarray}

where initial few coefficients are as follows  

$$\mathcal{A}_0(j) = -2\,\psi^{(0)}(j),$$

$$\mathcal{A}_1(j) = \frac{1}{6}\!\left[\pi^2 + 12\left(\psi^{(0)}(j)\right)^2 - 6\,\psi^{(1)}(j)\right],$$

$$\mathcal{A}_2(j) = -\frac{1}{3}\!\left[\pi^2\,\psi^{(0)}(j) + 4\left(\psi^{(0)}(j)\right)^3 - 6\,\psi^{(0)}(j)\,\psi^{(1)}(j) + \psi^{(2)}(j)\right]. \label{eq:app_coeffs}$$

Which further can be written in closed form in terms of Bell polynomias as
\begin{eqnarray}
  \frac{\Gamma(1-\eta-j)}{\Gamma(\eta+j)} = \frac{(-1)^j}{\left[(j-1)!\right]^2} \left[ \frac{1}{\eta} + \sum_{m=0}^{\infty} \frac{(-1)^m}{m!} B_{m+1}\!\!\left( 2\psi^{(0)}(j),\, -2\psi^{(1)}(j)+\frac{\pi^2}{3},\, 2\psi^{(2)}(j),\,\ldots \right)\eta^m \right], \nonumber\\
  \label{gammaf}  
\end{eqnarray}

Substituting Eq.~\eqref{gammaf} into Eq.~\eqref{finaleq} and isolating the $1/\eta$ pole from the finite remainder produces precisely the decomposition

$$x\mathcal{F}^{(1)}(x,k_\perp) = xf^{(1)}_g|_{\text{div}} + xf^{(1)}_g|_{\text{fin}},$$

written in Eqs.~\eqref{eq:div_part} and~\eqref{eq:fin_part} of the main text, from which the divergent and finite contributions are evaluated separately in Sec.~\ref{sec:level2}. The vanishing of the divergent piece confirms that no genuine ultraviolet divergence is present.

\bibliography{ref.bib}%
\end{document}